\documentclass[journal,10pt]{IEEEtran}

\usepackage{cite}
\usepackage{graphicx}
\usepackage[cmex10]{amsmath}
\usepackage{amsthm}
\usepackage{amsfonts}
\usepackage{caption}
\usepackage{epstopdf}
\usepackage{amssymb}
\usepackage{float}
\usepackage[export]{adjustbox}

\usepackage{color,soul}
\usepackage{subfig}
\usepackage{csquotes}
\usepackage{enumitem}
\usepackage{adjustbox,lipsum}
\usepackage{tabularx,booktabs}
\usepackage{array}
\newcolumntype{L}[1]{>{\raggedright\let\newline\\\arraybackslash\hspace{0pt}}m{#1}}
\newcolumntype{C}[1]{>{\centering\let\newline\\\arraybackslash\hspace{0pt}}m{#1}}
\newcolumntype{R}[1]{>{\raggedleft\let\newline\\\arraybackslash\hspace{0pt}}m{#1}}

\begin{document}
\title{Beamwidth Selection for a Uniform Planar Array (UPA) Using RT-ICM mmWave Clusters}
\author{Yavuz Yaman$^1$, \IEEEmembership{Member, IEEE}, and Predrag Spasojevic$^2$, \IEEEmembership{Senior Member, IEEE}
\thanks{$^1$The author is with Qualcomm Corporate R\&D, Bridgewater, NJ 08807 USA (e-mail: yyaman@qti. qualcomm.com). 

$^2$The author is with WINLAB, Department of Electrical and Computer Engineering, Rutgers University, Piscataway, NJ 08854 USA (e-mail: spasojev@winlab.rutgers.edu).

The work is completed when the authors are with Department of Electrical and Computer Engineering, Rutgers University, Piscataway, NJ 08854 USA.}}

\maketitle

\begin{abstract}
\boldmath
Beamforming is the primary technology to overcome the high path loss in millimeter-wave (mmWave) channels. Hence, performance improvement needs knowledge and control of the spatial domain. In particular, antenna structure and radiation parameters affect the beamforming performance in mmWave communications systems. For example, in contrast to common belief, narrow beamwidth may result in degraded beamforming performance. In order to address the impairments such as beam misalignments, outage loss, tracking inability, blockage, etc., an optimum value of the beamwidth must be determined. In our previous paper, assuming a communication system that creates a beam per cluster, we theoretically investigated the beamwidth and received power relation in the cluster level mmWave channels. We used uniform linear array (ULA) antenna in our analysis. In this paper, we revisit the analysis and update the expressions for the scenario where we use rectangular uniform planar array (R-UPA) antenna. Rectangular beam model is considered to approximate the main lobe pattern of the antenna. For the channel, we derive beamwidth-dependent extracted power expressions for two intra- cluster channel models, IEEE 802.11ad and our previous work based on ray-tracing (RT-ICM). Combining antenna and channel gains, in case of the perfect alignment, we confirm that the optimum beamwidth converges zero. Performing asymptotic analysis of the received power, we give the formulation and insights that the practical nonzero beamwidth values can be achieved although sacrificing subtle from the maximum received power. Our analysis shows that to reach $95\%$ of the maximum power for a typical indoor mmWave cluster, a practical beamwidth of $3.5^{\circ}$ is enough. We also investigate the channel affect where we compare the two channel models and show that although the difference in beamwidth is less than $1^{\circ}$, the difference in required number of elements can reach to $50$. Finally, our analysis results show that there is a $13$ dB increase in the maximum theoretical received power when UPA is used over ULA for the same cluster. We show that a $8\times8$ UPA can reach $50\%$ of that maximum received power while the received power is still $10$ dB larger than the ULA scenario. In the simulation section, we show that the expressions given by the analysis match to the simulated results.

\end{abstract}

\section{Introduction}\label{intro}

\IEEEPARstart{M}{illimeter}-wave (mmWave) communication has several advantages over the current wireless bands such as higher throughput, lower latency, reduced interference, and increasing network coordination ability. Nevertheless, high path loss is the significant drawback of mmWave channels. To overcome, beamforming is proposed as a substantial solution with the availability of large array usage in a small-scale area. On the other hand, due to the sparse nature of mmWave channels, clusters are generally spatially-separated \cite{Maltsev_conference}. That further allows creating a beam for each cluster, both in the transmitter and the receiver end which, in turn, yields increased performance in multi-input-multi-output (MIMO) and massive MIMO applications \cite{11ay}. Significant contribution is published for the receiver processing aspect of beamforming, including optimum transmitter and receiver design \cite{Ayach} with array antennas and beamforming protocols \cite{Yaman}. However, maximizing the beamforming efficiency can be challenging due to the misalignments, weak tracking ability, blockages, outage loss, etc. which requires channel knowledge in the angular domain. Specifically, beams with non-optimized beamwidths may increase inter-beam interference and wasted energy, i.e. outage rate, or even cause a link failure easily when combined with beam misalignment. As a result, while the requirement of the accuracy on the beam alignment to the cluster angle of arrival (AoA) is unquestionable, selecting an appropriate beamwidth is also essential in the mmWave system networks. 
%In contrast to common belief, narrow beamwidth,...

Several measurements are already conducted in mmWave communications and prove that beamwidth has a critical effect on the channel parameters. In \cite{Manabe, Williamson,Kim}, antenna directivity (indirectly beamwidth) dependency to the delay and angle spread of the link is investigated at 28, 38 and 60 GHz. %These experiments show that delay and angle spread are proportional to beamwidth both for LOS and NLOS cases. 
\cite{Rappaport_aoa, Lee} conduct some outdoor experiments at 28 and 38 GHz with different beamwidth antennas and measure the incurred path loss. \cite{Rajagopal} provides similar outdoor LOS and NLOS tests and collect data of captured energy (received power) for several beamwidth values at 28 and 40 GHz. %Both experiments show that wider beams have better performance, i.e. capture more energy and experience less path loss. 
In \cite{Dogan}, optimum beamwidth is measured in case of blockage occurs within the channel where wider beamwidths are provided based on beam expansion. %This is simply due to reason that beamforming of the channel, spatially, filters some information out too. 
While the nonnegligible effect of the beamwidth on mmWave communications is demonstrated with several other measurement results, on the other hand, very few beamwidth analyses on the performance metrics are proposed so far. In \cite{Va}, it is shown that there is an optimal non-zero beamwidth (around $5^{\circ}$) that maximizes the coherence time of the time-varying vehicular channel at 60 GHz. In \cite{Akoum}, analysis results show that $10^{\circ}$ beamwidth has better coverage, less interference compared to $30^{\circ}$ for mmWave cellular networks. A more related work \cite{Vakilian} studies AoA estimation error effects on bit-error-rate (BER) with different beamwidths for the clustered channel model. %However, the focus is on the AoA estimation errors and the optimum beamwidth problem is still not addressed. Furthermore, a reconfigurable antenna is taken into account in the analysis which is impractical in mmWave systems. 
A detailed analysis of the link between the channel angular dispersion and the antenna structure is given in \cite{Yang}. However, the channel is simply assumed to be Rician and no clustering approach is adopted as generally seen in mmWave channels. 

In \cite{Yaman3}, we provided an analytical framework for the optimum beamwidth that maximizes the received power for indoor mmWave clusters, in the case of both perfect alignment and misalignment where we used uniform linear array (ULA) antenna at the receiver. In this paper, we do the similar analysis except that the receiver antenna in the analysis is updated to uniform planar array (UPA) which is planned to be more preferred for mmWave devices. We also derive the formulas only for perfect alignment of the beam to the cluster angle profile. We first give the relation between beamwidth and the captured power from the cluster. To do so, we use two different intra-cluster channel models, IEEE 802.11ad \cite{11ad} and our previous work, RT-ICM \cite{Yaman2}. Then, we combine it with the antenna gain of the beam steered to the cluster AoA and provide an overall received power and beamwidth relation. Here, we derive the directivity (and gain) for different scenarios where the elevation angles are fixed as the cluster channel models are provided only at the azimuth domain. We give two main lobe beam models but consider only rectangular beam model in the formulation to approximate the main lobe array pattern. We show that for the perfect alignment, the optimum beamwidth is at zero, while the theoretical maximum received power approaches to a constant, similar to the result in \cite{Yaman3} . Then, we give practical limits of the optimum beamwidth with the relation to the number of elements such that sacrificing from the maximum received power in the order of tenths can reduce the required number of antenna elements significantly. We evaluate the performance of the analysis by comparing the analytical results with simulations for an indoor mmWave cluster. Finally, we give the comparison of the UPA and ULA usage in terms of the received power, beamwidth and the number of elements required. The work we propose in this paper will give insights to the optimum antenna array design in MIMO applications for future mmWave systems network. 

The rest of the paper is organized as follows. In Sec. \ref{system_model}, the overall problem is defined analytically and RT-ICM is summarized. In Sec. \ref{Antenna}, the relation between the beamwidth and the antenna structure is given. Sec. \ref{problemformulation} gives the problem formulation for two cluster models and provides simplified expressions for the optimum beamwidth. Performance evaluations of the extracted expressions are analyzed in Sec. \ref{performanceevaluation}. Finally, Sec. \ref{conclusion} concludes the paper.

\section{System and Channel Model} \label{system_model}

The optimum beamwidth problem has two sides; while decreasing beamwidth and increasing directivity of an array beam, (1) antenna gain increases, (2) captured energy from the cluster channel decreases. Hence, received power on the antenna terminals directly depends on the beamwidth. %In this paper, we investigate the theoretical limits of the maximum received power from a cluster while beamwidth approaches to zero. 

\begin{figure}[t]
\centering
\includegraphics[scale=0.48]{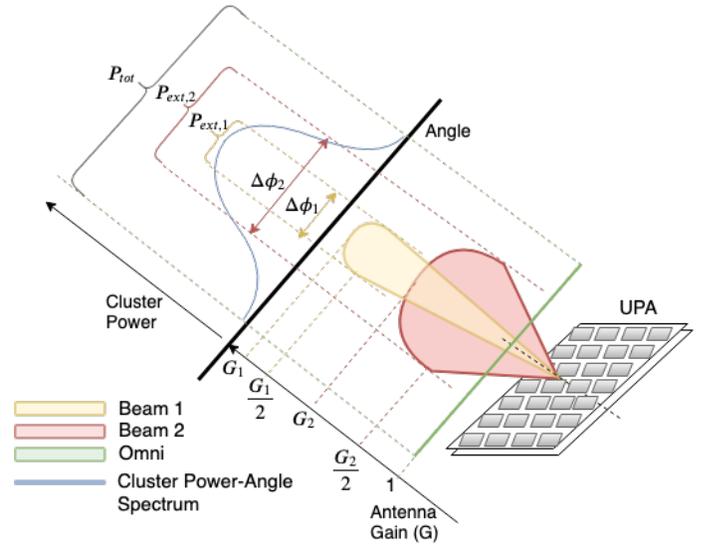} \\
\caption{Visualization of the optimum beamwidth problem at the receiver}\label{beamwidth}
\end{figure}
Received power on the antenna terminals is given in \cite{Orfanidis} as
\begin{equation}
P_R= \mathcal{P}_{inc}\frac{G\lambda^2}{4\pi} \label{PR}
\end{equation} 
where $\mathcal{P}_{inc}$ is the power density per \textit{area} front of the antenna; $G$ is the receiver antenna gain at maximum direction and $\lambda$ is the wavelength. Note that when an omnidirectional antenna is employed at the receiver, whose gain is $1$ at all directions, received power equals the available total cluster power. Then, the available power in front of the antenna can be given as

\begin{equation}
P_{tot}=\mathcal{P}_{inc}\frac{\lambda^2}{4\pi} \label{Ptot}
\end{equation}

Let $P_{ext}\leq P_{tot}$ is the extracted power from the cluster by a directional antenna. The equality holds whenever the beamwidth of the antenna covers entire cluster spatially\footnote{In this paper, we assume antenna beam models whose gain is 0 outside the beamwidth. That structure is discussed in Sec. \ref{Antenna}.}. Then, the received power given in Eq. (\ref{PR}) can be represented as a function of half-power beamwidth ($\Delta\phi$) and can be updated as
\begin{equation}
P_R(\Delta\phi)=G(\Delta\phi)P_{ext}(\Delta\phi) \label{PR_deltaphi}
\end{equation}

In Fig. \ref{beamwidth}, an example diagram of the discussion is illustrated with a comparison of two beams created by a UPA and steered towards a cluster AoA.  
%The effect of inverse relation between $\Delta\phi$ and $G_R$, and $P_{ext}$ can be seen from Eq. (\ref{PR_deltaphi}) immediately, as discussed in Sec. \ref{system_model}. In the next two sections, we give expressions for $G_R(\Delta\phi)$ and $P_{ext}(\Delta\phi)$.

Analysis of the problem requires the knowledge of spatial representation of the intra-cluster channel. However, while phased array antennas are well-studied in the literature and allow us to derive antenna gain-beamwidth relation, on the other hand, intra-cluster angular behavior of the mmWave channels is still not understood very well. In 3GPP channel model \cite{3GPP}, angular distribution of cluster power is simply modeled with a fixed number of rays with equal power levels. In 60 GHz WLAN standards IEEE 802.11ad \cite{11ad} and IEEE 802.11ay \cite{11ay}, a more intuitive model is adopted based on the measurements such that the power angular spectrum is distributed normally with $N(0, \sigma)$ where $\sigma=5$ for conference room and cubicle environments and $\sigma=10$ for living room channel models. Considering the site-specific nature of the mmWave channels, these models are likely to fail for different type of environments. %Hence, we introduced a simple ray tracing intra-cluster model for mmWave channels in our previous work \cite{Yaman2} that gives accurate results for any stationary environments. %In the next subsections, we summarize the model first and then express the research statement analytically by exploiting the adopted channel model.
	
	\subsection{Ray Tracing based Intra-Cluster Channel Model \cite{Yaman2}} \label{ICM}

\begin{figure}[t]
\centering
\includegraphics[scale=0.5]{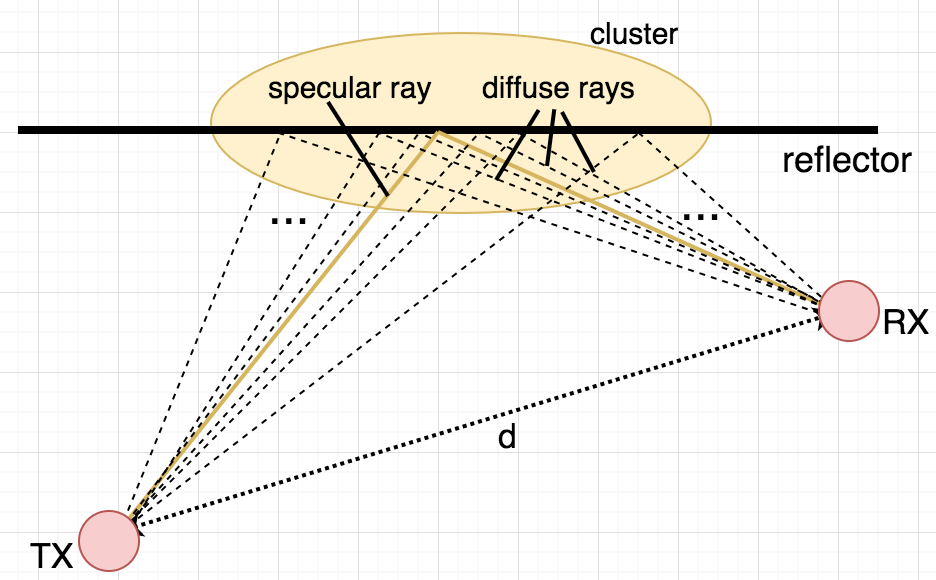} \\
\caption{First-order reflection cluster model of the RT-ICM}\label{clusterdef}
\end{figure}

In \cite{Yaman2}, we introduce a mmWave intra-cluster model based on ray-tracing (RT-ICM) that takes only first-order reflections into account. The general first-order reflection cluster definition of the model is illustrated in Fig. \ref{clusterdef}. It outputs the angular and temporal power distribution within the cluster and can be used for both indoor and outdoor mmWave systems in any type of stationary environments. The output theoretical baseband cluster impulse response of RT-ICM is given by \cite{Yaman2}
\begin{equation}
\begin{split}
c(t,\omega)&=a_{s} e^{j\varphi_{s}}\delta (t-t_{s})\delta (\omega-\Omega_s)\\
&+\sum_{k=0}^{N_r-1} a_k e^{j\varphi_k} \delta (t-t_{s}-\tau_k) \delta (\omega-\Omega_s-\alpha_k) \label{CIR_theo}
\end{split}
\end{equation}
where, $t$ and $\omega$ are the reference ToA and AoA variables; $a_{s}$, $\varphi_{s}$, $t_{s}$, and $\Omega_s$ are the amplitude, phase, ToA and the AoA of the specular ray; $a_{k}$, $\varphi_{k}$, $\tau_{k}$, $\alpha_{k}$ are amplitude, phase, delay, offset AoA of the $k$-th ray, respectively. $\delta (.)$ is  Dirac delta function and $N_r$ is the number of diffuse rays. 

\begin{figure}[t]
\centering
\includegraphics[scale=0.23]{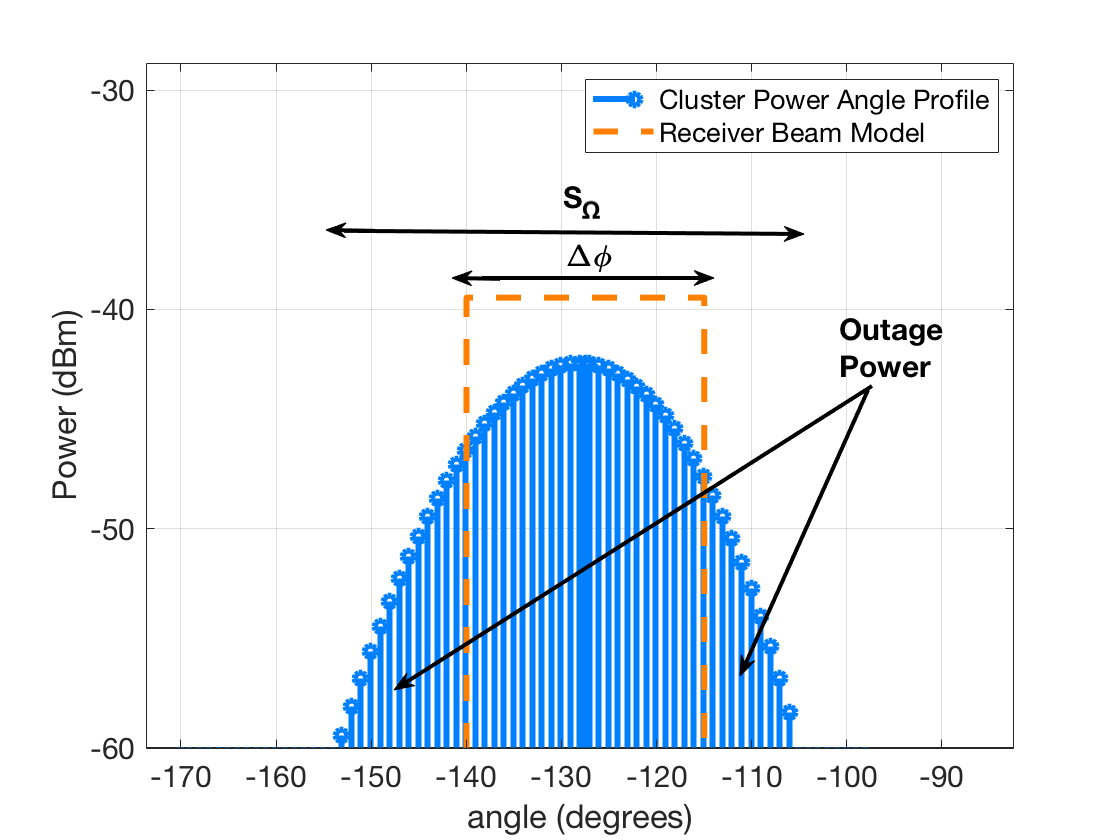} \\
\caption{An example cluster power angle profile of RT-ICM.}\label{outage_figure}
\end{figure}

Then, the total coherent power in the cluster is given as
\begin{equation}
P_{tot}=(a_{s})^2 + \frac{S_{\Omega}}{N_r} \sum_{k=0}^{N_r-1} (a_k)^2  \label{totalpower}
\end{equation}
where $S_{\Omega}=\alpha_{N_r-1}-\alpha_{0}$ is the supported angle spread (SAS) at the receiver and $S_{\Omega}/N_r$ term is inserted to compansate the digitization of the spatial domain. An example of a cluster angle profile output of RT-ICM is displayed in Fig. \ref{outage_figure} for $N_r=75$ and $S_{\Omega}=75^{\circ}$.

\section{Antenna Structure and Gain} \label{Antenna}

\begin{figure}[t]
\centering
\includegraphics[scale=0.30]{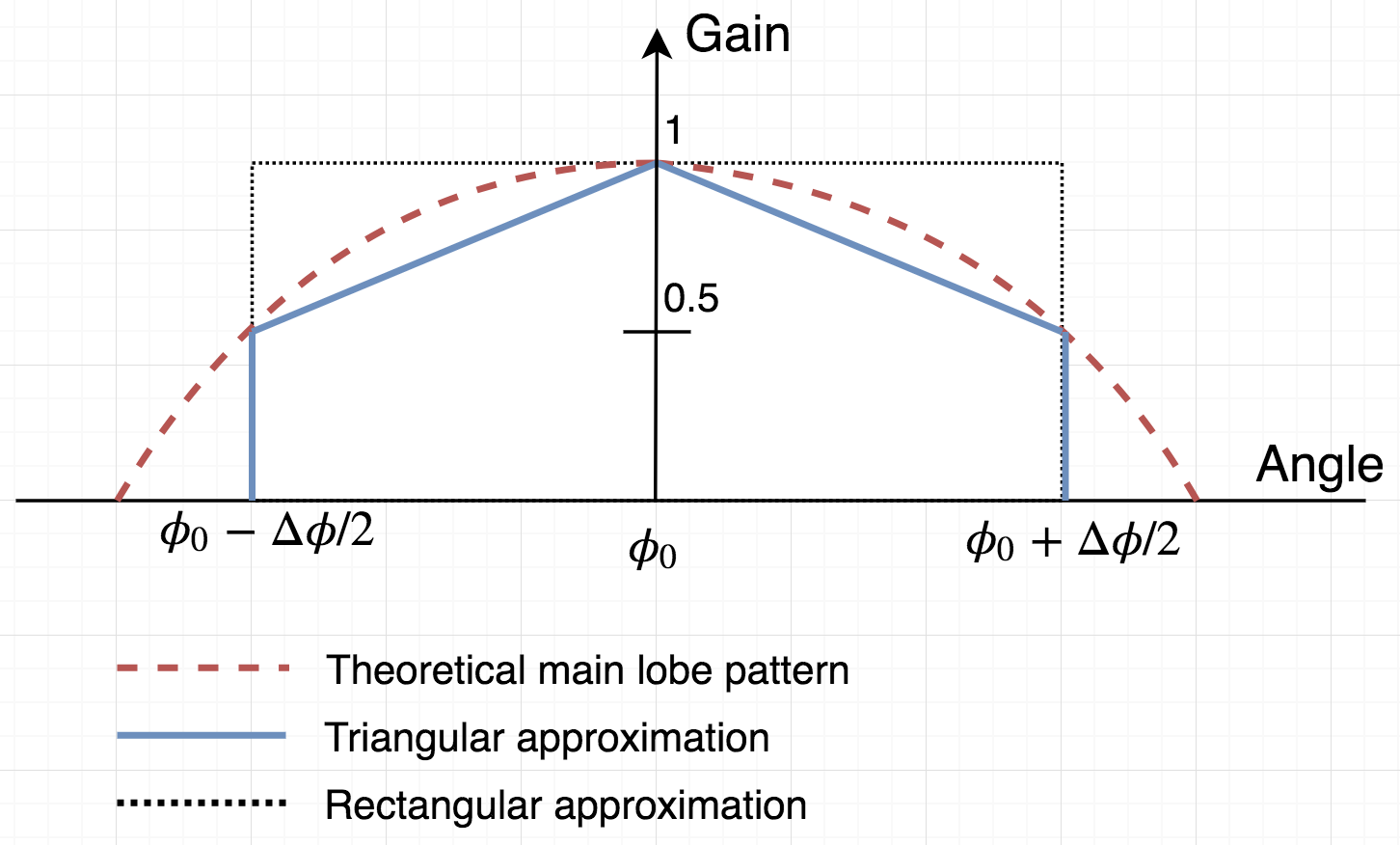} \\
\caption{Antenna pattern models considered in the paper.}\label{patternapprox}
\end{figure}

As Eq. (\ref{PR_deltaphi}) suggests, antenna gain is the counterpart of the captured cluster power in the equation for a certain beamwidth. In \cite{Yaman3}, we simply assumed a well-known uniform linear array (ULA) design where the spacings between the elements are equal and provide the relationships between antenna gain, beamwidth, number of elements and scan angle for a ULA. In this paper, we adopt the rectangular uniform planar array (R-UPA) antenna model as it is a more realistic and practical design in mmWave devices. In that aspect, we first derive the relation between the antenna gain and beamwidth, and then use the result to analyze the maximum received power by combining the extracted power from the cluster. 

\subsection{Beam Pattern Model} \label{beampatternmodel}
Two models are used for the beam pattern; a rectangular window and a triangular window, both are seen in Fig. \ref{patternapprox} for a steering (scan) angle of $\phi_0$. Both approximations ignore the sidelobes; thereby modeling only the main lobe. Expressions of the shown functions for rectangular and triangular model, respectively, are,
\begin{align}
W_{R}= &
\begin{cases} \label{WR}
1,  & \phi_0-\Delta\phi/2 > \phi > \phi_0+\Delta\phi/2 \\
0, & otherwise
\end{cases}\\ 
W_{T}= &
\begin{cases} \label{WT}
1-\frac{|\phi-\phi_0|}{\Delta\phi},  & \phi_0-\Delta\phi/2 > \phi > \phi_0+\Delta\phi/2 \\
0, & otherwise
\end{cases}
\end{align}
 
\subsection{Directivity of RUPA}

\begin{figure}[t]
\centering
\includegraphics[scale=0.55]{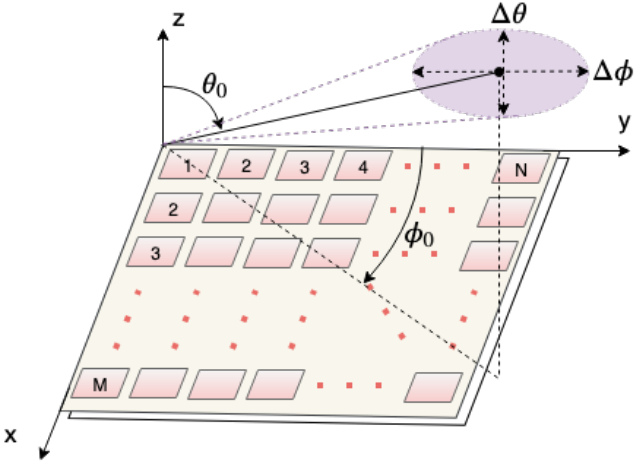} \\
\caption{$(N \times M)$-element rectangular UPA diagram with the scan angles and beamwidths illustration.}\label{upamodel}
\end{figure}

A diagram of a $M \times N$-element rectangular UPA is given in Fig. \ref{upamodel}. ($\theta_0$, $\phi_0$) and ($\Delta\theta$, $\Delta\phi$) pairs are the elevation and azimuth scan angles and the beamwidths of the beam, respectively. The goal in this subsection is to find an expression for the array directivity in terms of the azimuth beamwidth. To simplify the analysis, we assume that the x and y direction inter-element spacings are both set to $\lambda/2$. 

For large $N$ and $M$, the directivity of a rectangular UPA is given in \cite{Balanis} as $D=\pi \cos \theta_0 D_x D_y$ where $D_x$ and $D_y$ are the directivities of the x-axis and the y-axis ULAs, respectively. Assuming the uniform excitation of all antenna elements, directivity of the x- and y-axis ULAs at broadside equals the number of elements on the corresponding axes, i.e. $D_x=M$ and $D_y=N$ \cite{Hansen}. Then, the UPA directivity becomes
\begin{equation}
D=\pi \cos \theta_0 NM  \label{directivity}
\end{equation}

Further, using the beamwidth and the number of elements relation of a uniformly excited ULA with $\lambda/2$ spacing \cite{Balanis,Orfanidis,Hansen} where $\Delta\phi_x=101.5/M$ and $\Delta\phi_y=101.5/N$ are the beamwidths of the x- and the y-axis ULAs, respectively, Eq. (\ref{directivity}) is updated as

 \begin{equation}
D=\pi \cos \theta_0 \frac{101.5^2}{\Delta\phi_x\Delta\phi_y}  \label{directivity2}
\end{equation}

On the other hand, the azimuth and the elevation beamwidths of a large rectangular UPA for arbitrary $\theta_0$ and $\phi_0$ are given in \cite{Balanis}, respectively, as,

\begin{equation}
\Delta\theta=\sqrt{\frac{1}{\cos^2 \theta_0 (\Delta\phi_x^{-2} \cos^2\phi_0 + \Delta\phi_y^{-2} \sin^2\phi_0 )}} \label{deltatheta}
\end{equation}
\begin{equation}
\Delta\phi=\sqrt{\frac{1}{\Delta\phi_x^{-2} \sin^2\phi_0 + \Delta\phi_y^{-2} \cos^2\phi_0}} \label{deltaphi}
\end{equation} 

After combination of Eq. (\ref{deltatheta}) and (\ref{deltaphi}) and solving for $\Delta\phi_x$, we get
\begin{equation}
\Delta\phi_x=\frac{\Delta\theta \cos \theta_0 \Delta\phi_y \Delta\phi}{\sqrt{\Delta\phi^2 (\Delta\phi_y^2 - \Delta\theta^2 \cos^2 \theta_0) + \Delta\theta^2 \cos^2 \theta_0\Delta\phi_y^2 }} \label{deltaphix}
\end{equation}

Plugging $\Delta\phi_x$ into Eq. (\ref{directivity2}),
\begin{equation}
D= \frac{101.5^2\pi \sqrt{\Delta\phi^2 (\Delta\phi_y^2 - \Delta\theta^2 \cos^2 \theta_0) + \Delta\theta^2 \cos^2 \theta_0\Delta\phi_y^2}}{\Delta\theta \Delta\phi \Delta\phi_y^2} \label{directivity_comp}
\end{equation}

As expected, directivity depends also on elevation scan angle ($\theta_0$) and beamwidth ($\Delta\theta$) as well as y-axis ULA beamwidth at broadside ($\Delta\phi_y$). In order to be able to work with the channel side power angle spectrum where only the azimuth domain information is available, in the next subsection, we will pick reasonable selections for those parameters by discussing the theoretical constraints. 

\subsection{Selection of Angular Parameters}

\subsubsection{Constraint 1} Immediate first constraint is already imposed within the formulation of the UPA directivity. The large array assumption puts a limitation on $\Delta\phi_y$. Considering the large array limitation as $N \geq 7$ \cite{Balanis},  the first constraint is setup as: $\Delta\phi_y \leq 101.5/7= 14.5^{\circ}$.

\subsubsection{Constraint 2} Apparently, all beamwidth and scan angle values should be positive and real. From Eq. (\ref{deltaphix}), to provide a real beamwidth value, the square root term should be real. To simplify the analysis further, we tighten the limitation and set the constraint as: $\Delta\phi_y^2 - \Delta\theta^2 \cos^2 \theta_0 \geq 0$. %Even tighter, $\Delta\phi_y^2 \geq \Delta\theta^2$.

\subsubsection{Constraint 3} Finally, to ensure the aligned communication between two devices located at approximately the same height, we consider another constraint as: $\theta_0+\Delta\theta/2 \approx 90^{\circ}$. The meaning of this constraint can be visually inspected from Fig. \ref{upamodel}.

Based on these constraints, we derive the directivity expressions for a few different selections of the three parameters, $\theta_0, \Delta\theta, \Delta\phi_y$. In Table \ref{candidate_selections}, 4 set of candidate parameter values are listed with their derived directivity formulas by plugging them into Eq. (\ref{directivity_comp}). Set $1$ and $2$ don't violate any constraint, however, constraint $2$ doesn't hold for Set $3$. Set $4$ violates constraints $2$ and $3$. 

\begin{table}[!t]
\centering
\caption{Some candidate selections of the parameters} 
\label{candidate_selections}
\begin{tabular}{|c||c|c|c||c|} 
\hline 
Set ID & $\Delta\phi_y$  & $\Delta\theta$ &  $\theta_0$ & Directivity ($D$)
  \\ \hline \hline \vspace{1mm}
1 & $14.5^{\circ}$ & $30^{\circ}$ & $75^{\circ}$ &  $\frac{20\pi\sqrt{84.68+\Delta\phi^2}}{\Delta\phi}$ 
\\  \hline \vspace{1mm}
2 & $14.5^{\circ}$ & $40^{\circ}$ & $70^{\circ}$ &   $\frac{5.91\pi\sqrt{1703+\Delta\phi^2}}{\Delta\phi}$ 
\\ \hline \vspace{1mm}
3 & $10.15^{\circ}$ & $40^{\circ}$ & $70^{\circ}$ &   $\frac{22.93\pi\sqrt{229-\Delta\phi^2}}{\Delta\phi}$ 
 \\  \hline \vspace{1mm}
4 & $10.15^{\circ}$ & $30^{\circ}$ & $60^{\circ}$ &   $\frac{45.9\pi\sqrt{190-\Delta\phi^2}}{\Delta\phi}$ 
\\ \hline
\end{tabular}
\end{table}

\begin{figure}[t]
\centering
\includegraphics[scale=0.24]{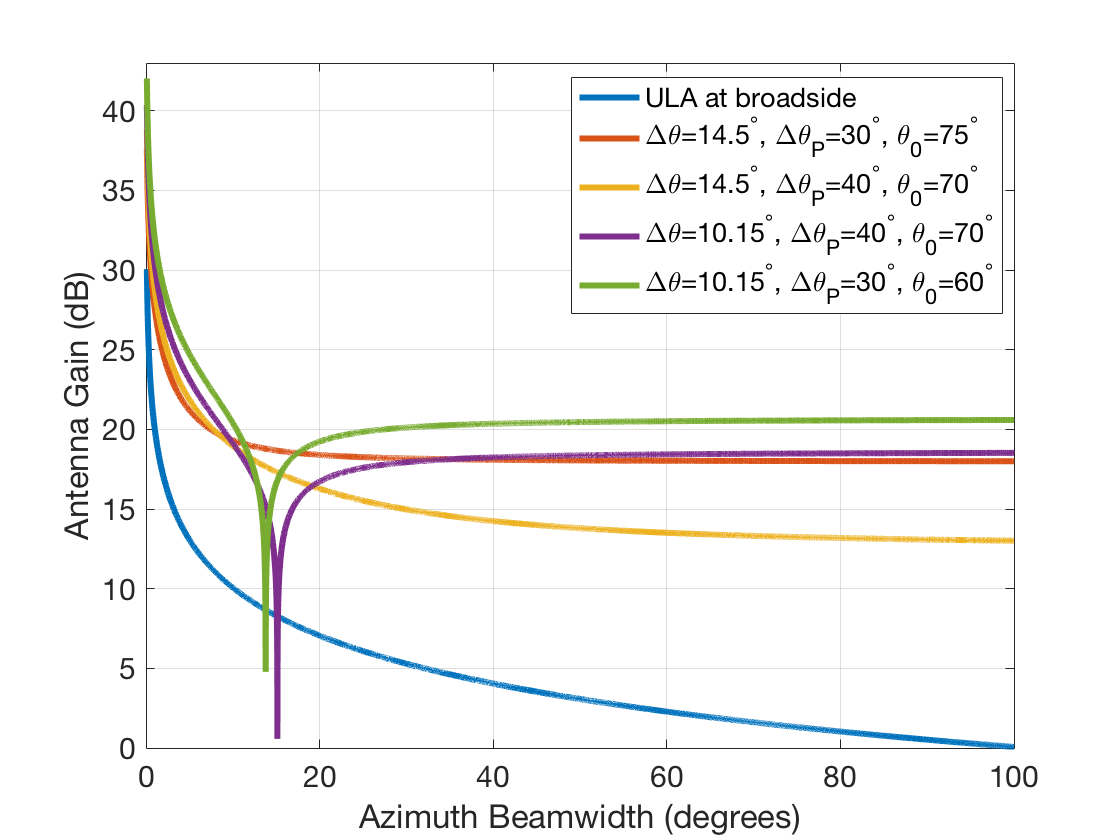} \\
\caption{Directivities of the selected parameter sets and the effect of the constraints.}\label{setdirectivities}
\end{figure}

Directivity and the azimuth beamwidth relation for these sets are plotted in Fig. \ref{setdirectivities}. The effect of the violation of constraint $2$ can be seen for Set $3$ and $4$ where the azimuth beamwidth is supported for a limited range.

Since the efficiency of phased array antennas are almost perfect \cite{Hansen}, we use antenna gain and directivity interchangeably, i.e. $G=D$.

\section{Problem Formulation} \label{problemformulation}

\subsection{Extracted Power from Channel} \label{extracted power}

For a perfect beam alignment to the channel, extracted power from IEEE 802.11ad intra-cluster Gaussian channel model was formulated in \cite{Yaman3} as

\begin{equation}
P_{ext}^{st}= P_{tot} \int_{\phi_{cl}-\Delta\phi/2}^{\phi_{cl}+\Delta\phi/2} W_R(\phi) \frac{1}{\sqrt{2\pi\sigma^2}}e^{-\frac{(\phi-\phi_{cl})^2}{2\sigma^2}} \text{d}\phi  \label{Pextst}
\end{equation}

where  $\phi_{cl}$ is the cluster AoA. In the equation, we used rectangular beam shape given in Eq. (\ref{WR}). 

For RT-ICM, we show in \cite{Yaman3} that the Gaussian approximation of the resultant power angle profile fits almost perfectly to the data. Then, the extracted power from RT-ICM is obtained by 

\begin{equation}
P_{ext}^{rt}=\int_{x-\Delta\phi/2}^{x+\Delta\phi/2} W_R(\phi)  g(\phi) \text{d}\phi \label{PextICM_gauss}
\end{equation} 

where $g(\phi)=ue^{(\phi-x)^2/v^2}$ is the Gaussian function that fits the data with the calculated variables $u$, $x$ and $v$\cite{Guo}.

\subsection{Received Power Problem Formulation}

As Eq. (\ref{PR_deltaphi}) suggests, the received power from a cluster is the multiplication of the antenna gain and the extracted power by the antenna. At this point, for antenna gain ($G=D$) we will pick a scenario from the Table \ref{candidate_selections} and derive the received power accordingly. In fact, since the optimized beamwidth falls into the region below $10^{\circ}-15^{\circ}$ \cite{Yaman3}, any of the candidate sets in the table could be used as their similar performance can be seen from Fig. \ref{setdirectivities}. In this paper, we select the Set 4 and use the provided directivity equation to formulate the received power.

On the other hand, from \cite{Papoulis}, the integration of a Gaussian can be defined with the error function (\textit{erf}). Then the extracted power for 802.11ad given in Eq. (\ref{Pextst}) becomes

\begin{equation}
P_{ext}^{st}= P_{tot}\text{erf} \left( \frac{\Delta\phi}{2\sqrt{2}\sigma} \right)
\end{equation}

Plugging the Set 4 antenna gain in Table \ref{candidate_selections} and $P_{ext}^{st}$ into Eq. (\ref{PR_deltaphi}), received power as a function of beamwidth can be given as

\begin{equation}
P_R=\frac{45.9P_{tot}\pi\sqrt{190-\Delta\phi^2}}{\Delta\phi}\text{erf} \left( \frac{\Delta\phi}{2\sqrt{2}\sigma} \right) \label{PR_st_perfect}
\end{equation}

\subsubsection{Maximum Received Power}
As the analysis shows in \cite{Yaman3}, optimum beamwidth that maximizes the received power in the case of perfect alignment converges to zero, i.e. $\Delta\phi_{opt} \to 0$. Hence, we perform an asymptotic analysis to get the theoretical maximum received power. Eq. (\ref{PR_st_perfect}) is in the $0/0$ indeterminate form for $\Delta\phi =0$. Applying L'Hopital rule, the maximum achievable received power is 
 \begin{equation}
 P_{max}^t=P_R(0)=\frac{793P_{tot}}{\sigma} \label{Pmax_st}
 \end{equation}
 Intermediate steps are omitted as it is similar to our previous work \cite{Yaman3}.
 
 \subsubsection{Optimum Practical Beamwidth} \label{11ad_opt_beam}
 Note that the maximum received power given in the Eq. (\ref{Pmax_st}) is theoretical. Achieving very small beamwidth requires impractically high number of antenna elements. However, we can keep the beamwidth in practical ranges while sacrificing subtle from the received power. 

Let $0<\eta\leq 1$ be the coefficient such that 
\begin{equation}
P_{\eta}^{t}=\eta P_{max}^{t} \label{Peta}
\end{equation} 
where $P_{\eta}^{t}$ is the $\eta$-percentile power of the $P_{max}^{t}$. Then, from Eq. (\ref{PR_st_perfect}),
\begin{equation}
\frac{144.2P_{tot}\sqrt{190-\Delta\phi^2}}{\Delta\phi } \text{erf} \left( \frac{\Delta\phi}{2\sqrt{2}\sigma} \right)=P_{\eta}^{st}=\eta \frac{793P_{tot}}{\sigma}
\end{equation}

Simplifying the equation and setting $\Delta\phi=\Delta\phi_{\eta}$, i.e. practical $\eta$-percentile beamwidth,
\begin{equation}
\frac{\Delta\phi_{\eta}}{\text{erf} \left( \frac{\Delta\phi_{\eta}}{2\sqrt{2}\sigma} \right) \sqrt{190-\Delta\phi_{\eta}^2}}  =  \frac{\sigma}{5.5\eta} \label{deltaerf}
\end{equation}

Compared to ULA case in \cite{Yaman3}, for $\sigma=5^{\circ}$, Eq. (\ref{Pmax_st}) produces $\approx 13$ dB more power. It is basically due to the higher antenna gain in UPA. 

For completeness, we give the RT-ICM counterpart expressions of Eq. (\ref{Pmax_st}) and Eq. (\ref{deltaerf}), respectively, as

\begin{equation}
 P_{max}^r=632.7\pi u \label{Pmax_rt}
\end{equation}
\begin{equation}
\frac{\Delta\phi_{\eta}}{\text{erf} \left( \frac{\Delta\phi_{\eta}}{2v} \right) \sqrt{190-\Delta\phi_{\eta}^2}}  =  \frac{v}{7.8\eta} \label{deltaerf_rt}
\end{equation}

\section{Performance Evaluation} \label{performanceevaluation}

\begin{figure}[t]
\centering
\includegraphics[scale=0.24]{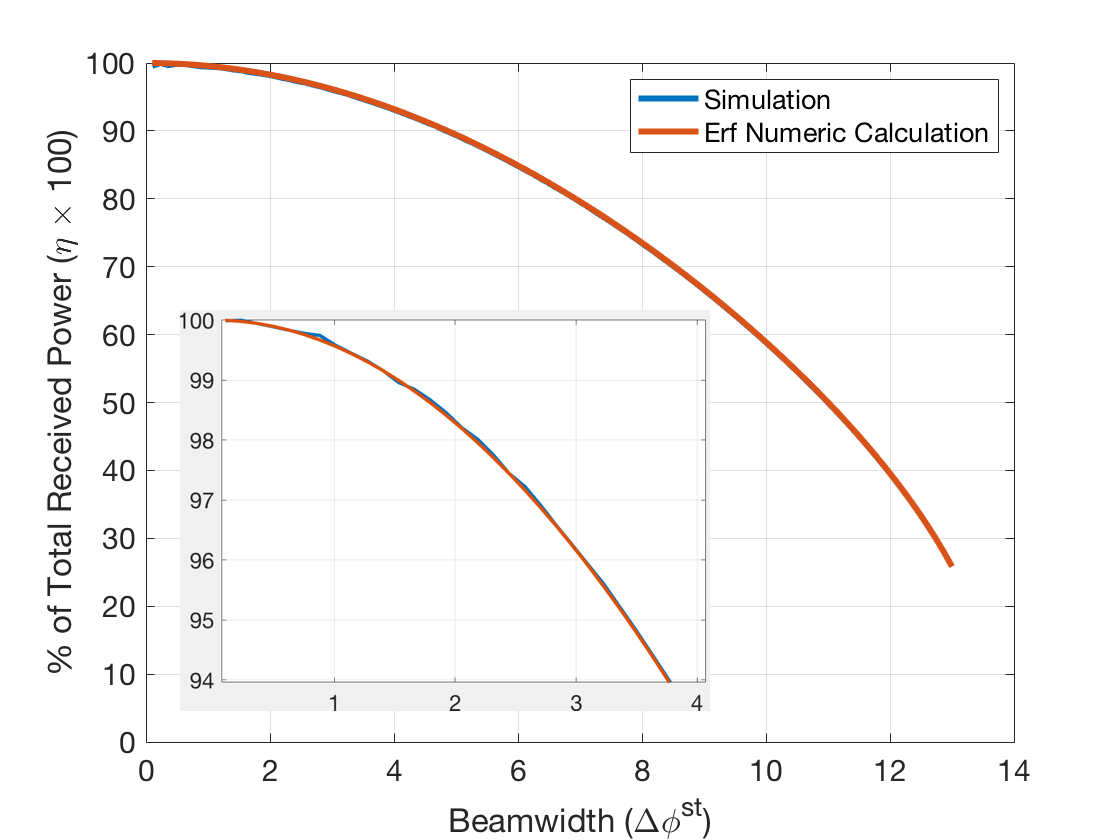} \\
\caption{Percentage of received power versus beamwidth when $\sigma=5$.}\label{st_UPA_percentile}
\end{figure}

In this section, we plot the performance of the derived expressions. In the first plot, we simulate the required beamwidth to obtain a percentile of the total achievable power and compare it with Eq. (\ref{deltaerf}) for $\sigma =5^{\circ}$. As seen from Fig. \ref{st_UPA_percentile}, the numerical plot of Eq. (\ref{deltaerf}) perfectly agrees with the simulation, as expected. The most important result of the plot is that while an impractical $0^{\circ}$ beamwidth is required to achieve maximum received power, a practical $\approx3.5^{\circ}$ beamwidth value captures $95\%$ of it. 

\begin{figure}[t]
\centering
\includegraphics[scale=0.24]{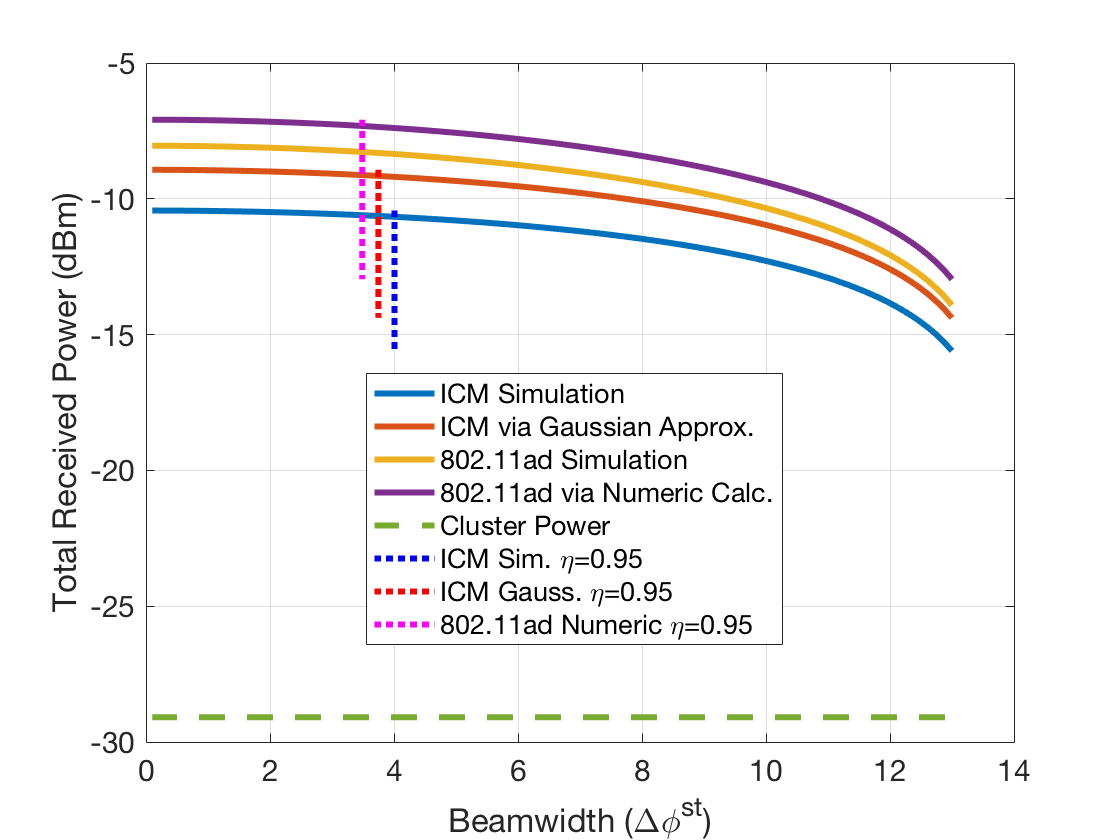} \\
\caption{Total received power versus beamwidth.}\label{TotalRecPow_bmW}
\end{figure}

\begin{figure}[t]
\centering
\includegraphics[scale=0.24]{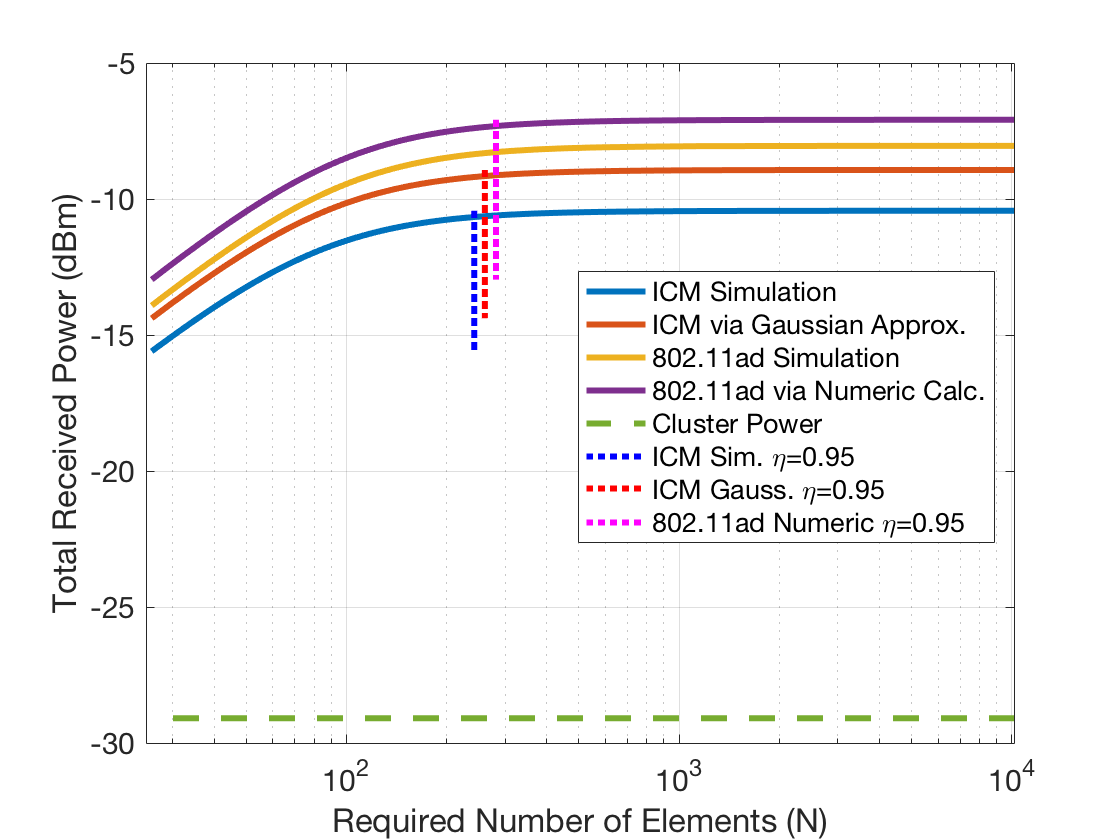} \\
\caption{Total received power versus number of elements.}\label{totrecpow}
\end{figure}

Next, in order to understand the channel effects, we run RT-ICM for a conference cluster \cite{Yaman2} and obtain the followings: $P_{tot}=-29.09$ dBm, $\phi_{cl} = 90^{\circ}$ and $S_{\Omega} = 72.2^{\circ}$. After fitting to the Gaussian function, $u = 6.434e-5$ and $v = 9.23$, which translates into $\sigma_r = \sqrt{2} v = 6.53^{\circ}$ in terms of Gaussian distribution. This corresponds to $1.53^{\circ}$ difference compared to IEEE 802.11ad model with $\sigma = 5^{\circ}$. In Fig. \ref{TotalRecPow_bmW}, we investigate the effect of the $\sigma$ difference by comparing IEEE 802.11ad and RT-ICM models as well as their $95\%$-percentile powers in terms of the beamwidth $\Delta\phi$. As seen, the difference in azimuth beamwidth can be assumed negligible ($<1^{\circ}$). In Fig. \ref{totrecpow}, we also illustrate the required number of elements ($N$) counterpart of the comparison. Although the difference in $N$ also seems subtle between the models and/or their approximations due to the logarithmic display, it can reach $30-50$ elements difference which can be critical in the hardware-limited applications. 

Finally, we compare the ULA vs UPA for the same cluster. As discussed in the previous section, approximately $13$ dB difference is visible which basically comes from the antenna gain. Note that to reach $95\%$ of the ULA's maximum power, one needs $\Delta\phi = 5.6^{\circ}$ which can be generated using $N = 19$ elements in ULA. However, $95\%$-percentile power of UPA case can be obtained by $\Delta\phi = 3.4^{\circ}$ which translates into $N = 290$ elements. On the other hand, even for $50\%$-percentile in UPA case, the received power is still 10 times larger ($10.2$ dB) than ULA usage. That power can be obtained in UPA with $11^{\circ}$ with using $N = 60$ elements. A typical practical UPA implementation can be $8\times8$ array structure.

\begin{figure}[t]
\centering
\includegraphics[scale=0.24]{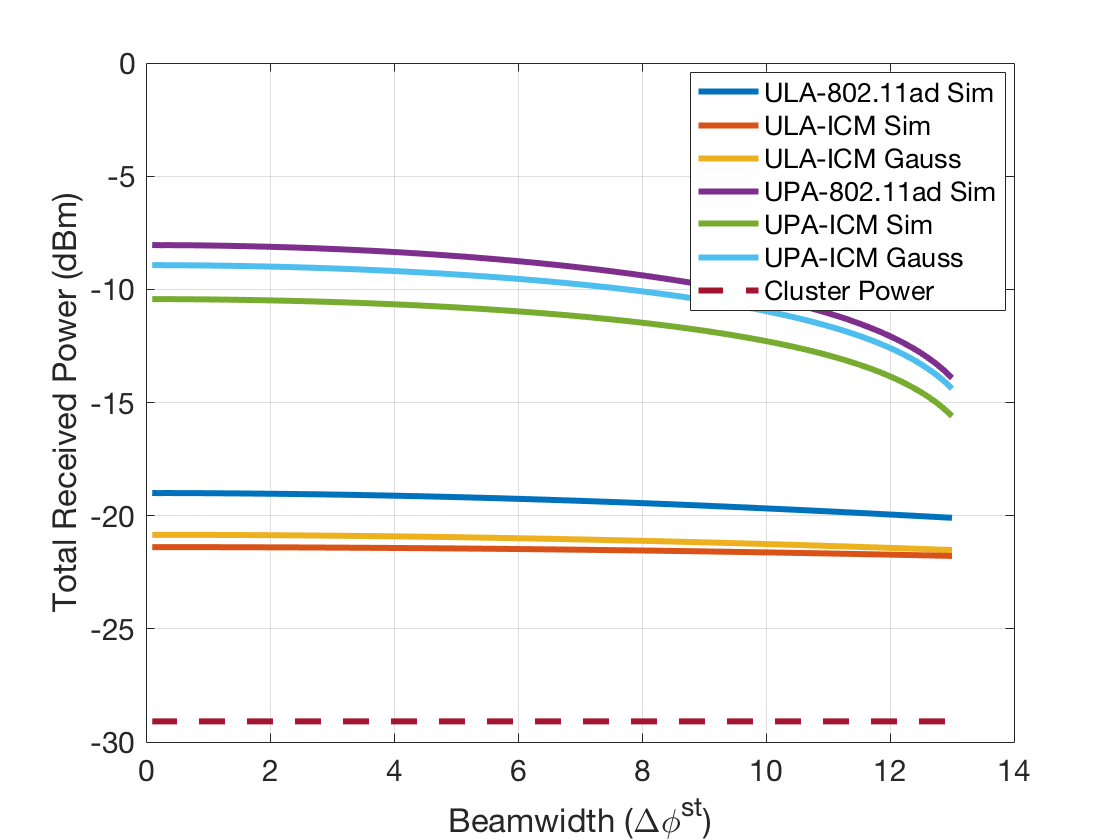} \\
\caption{ULA and UPA comparison for the same case study.}\label{ULAvsUPA}
\end{figure}

\section{Conclusion} \label{conclusion}

In this paper, we provide an analytical framework for the optimum beamwidth that maximizes the received power for indoor mmWave clusters. In the analysis, we consider the rectangular uniform planar array (R-UPA) antenna for the antenna gain and beamwidth relation. We use rectangular beam model to approximate the main lobe array pattern. Expressions that relate the beamwidth and captured cluster channel gain are provided for two intra-cluster model, IEEE 802.11ad and Ray Tracing based Intra-Cluster Model (RT-ICM) and the optimization problem is introduced by combining the antenna gain. Perfect alignment scenarios are studied. We show that the theoretical maximum received power converges to a constant while optimum beamwidth approaches to zero. We then provide equations that would result in practical beamwidth values while sacrificing from the maximum received power in the order of tenths percentage. We evaluate the performance of the analysis by comparing the analytical results with simulations for an indoor mmWave cluster. Finally, we compare the received power and beamwidth relation of ULA and UPA usages where we also translate it into the required number of elements. The work we propose in this paper will give insights to the optimum antenna array design in MIMO applications for future mmWave systems.

%\appendices
%\section{Maximization of $P_R$ in case of Misalignment} \label{maxPRmisalign}
%\subsection{First Derivative} 
%\section{Maximization of $P_R$ for Perfect Alignment} \label{802.11ad}
%\subsection{Maximum Received Power Derivation}

%\subsection{Optimum Beamwidth Approximation}

\section*{Acknowledgment}
The authors would like to thank Prof. S. Orfanidis for many helpful discussions and his contributions to the antenna theory. 

\bibliographystyle{IEEEtran}

%Authors
%\newpage
\begin{IEEEbiographynophoto}{Yavuz Yaman}
received the B.S degree from the School of Engineering, Istanbul University, in 2011; and M.S. and Ph.D. degrees in electrical and computer engineering from Rutgers University, Piscataway, NJ, in 2014 and 2020, respectively. He is currently a Senior Systems Engineer with Qualcomm Technologies, Inc., USA. His research interests include channel modelling, beamforming, channel estimation, antenna propagations and phased antenna arrays.
\end{IEEEbiographynophoto}

\begin{IEEEbiographynophoto}{Predrag Spasojevic}
received the Diploma of Engineering degree from
the School of Electrical Engineering, University of Sarajevo, in 1990;
and M.S. and Ph.D. degrees in electrical engineering from Texas A\&M
University, College Station, Texas, in 1992 and 1999, respectively.
From 2000 to 2001, he was with WINLAB, Electrical and Computer
Engineering Department, Rutgers University, Piscataway, NJ, as a
Lucent Postdoctoral Fellow. He is currently  Associate Professor in
the Department of Electrical and Computer Engineering, Rutgers
University. Since 2001 he is a member of the WINLAB research faculty.
 From 2017 to 2018 he was a Senior Research Fellow with
Oak Ridge Associated Universities working at the Army Research Lab, Adelphi, MD.
 His research interests are in the general areas of
communication and information theory, and signal processing.

Dr. Spasojevic was an Associate Editor of the IEEE Communications
Letters from 2002 to 2004 and served as a co-chair of the DIMACS
Princeton-Rutgers Seminar Series in Information Sciences and Systems
2003-2004. He served as a Technical Program Co-Chair for IEEE Radio
and Wireless Symposium in 2010. From 2008-2011 Predrag served as the
Publications Editor of IEEE Transactions of Information Theory.
\end{IEEEbiographynophoto}

%\newpage


\end{document}